\documentclass[12pt]{article}
\usepackage{fullpage}
\usepackage{mathtools}
\usepackage{array, color}
\usepackage{amsmath}
\usepackage{listings}
\usepackage{csvsimple, booktabs}
\usepackage{subfiles}
\usepackage{tikz}
\usepackage{longtable, caption}
\usetikzlibrary{automata, positioning, arrows}

\title{Top Score in Axelrod Tournament}
\author{Frederick M.\ Vincent \and Dashiell E.A.\ Fryer}
\date{Spring 2021}

\begin{document}
\maketitle
\begin{abstract}
The focus of the project will be an examination of obtaining the highest score in the Axelrod Tournament. The initial design of the highest score in the Axelrod Tournament consisted of looking at the Cooperation rates of the top strategies currently in the Axelrod Library. 

After creating an initial Finite State Machine strategy that utilized the Cooperation Rates of the top players; our ten-state FSM finished within the top 35 players out of 220 in the short run time strategies in the Axelrod Library. After a quick evolutionary algorithm, 50 generations, our original ten-state FSM was changed into an eight-state FSM, which finished within the top 5 of the short run time strategies in the Axelrod Library. 

This eight-state FSM was then evolved again using a full evolutionary algortihm process, which lasted 500 generations, where the eight-state FSM evolved into a another eight-state FSM which finishes first in the Axelrod Library among the short run time strategies and the full Axelrod Tournament. From that final FSM, two of the eight states are inaccessible, so the final FSM strategy is a six-state FSM that finishes first in the Full Axelrod Tournament, against the short run time strategies as well as the long run time strategies.
\end{abstract}

\section{Introduction}

The objective of this project was to obtain the highest score in the Axelrod Tournament. There have been a number of strategies created from various disciplines to compete to obtain the highest score. Throughout the last several decades, there have been numerous strategies developed in order to compete in the tournament, as well as different strategy types \cite{axelrodproject}. This paper will demonstrate how our team was able to achieve the highest score in the Axelrod Tournament.

We have developed a Finite State Machine strategy that has six states that has the highest score in the Axelrod Tournament. Our six-state strategy was developed from a ten-state strategy that was written out by hand, \emph{SecondPrac}, that ten-state strategy was then evolved using the evolutionary algorithm into an eight-state Finite State Machine, \emph{Fourthprac}. This eight-state FSM finished in the top 5 of the short run time strategies before another evolutionary process that led it to the top score. The final eight-state Finite State Machine that was produced had two inaccessible states, so the final strategy was a six-state FSM called \emph{EvolvedFSM6.}

\section{Axelrod Tournament}
The Axelrod Tournament has over 220 strategies ranging from Finite State Machines, to Artifical Neural Networks, to LookerUp Tables, as well as other strategies designed to play the Prisoner's Dilemma \cite{axelrodproject}. The top scores in the Axelrod Tournament are evolved versions of the LookerUp Table, the Evolved Neural Network, as well as an Evolved 16-state Finite State Machine. Each one of those strategies has been evolved, or trained to be able to maximize its score. There are other strategies that make up the rest of the field, to be able to compare how different strategies fair against one another. Many of the strategies that are in the library are deterministic, they follow a set of pre-determined guidelines to determine which moves will be played in a given situation. All of the strategies that are currently in the Axelrod Library can be found at the Axelrod Library's Github page \cite{axelrod-python}.

When initially running the Axelrod Tournament with all of the short run time strategies in the Library, \emph{EvolvedLookerUp2\_2\_2} finishes first, followed by \emph{Evolved FSM 16}, and \emph{Evolved HMM 5}. Each of these strategies have been evolved and their scores can be seen in Table~\ref{tab:table}. The top strategies are all different forms of evolved strategies in the Axelrod Tournament.

\begin{table}[ht]
\caption{Top Strategies}\label{tab:table}

\csvreader[
  longtable=llr,
  table head= 
    \toprule\bfseries Rank &\bfseries Name &\bfseries Median Score\\ \midrule\endhead
    \bottomrule\endfoot,
  late after line=\\,
  filter= \value{csvrow} < 15,
  before reading={\catcode`\#=12},
  after reading={\catcode`\#=6}
]{csvs/thesis.csv}{1=\Rank,2=\Name, 3=\Median_score}{\Rank & \Name & \Median_score}
\end{table}
{\addtocounter{table}{-1}} 

\section{Idea}
The main idea was to create a Finite State Machine to obtain the highest score. The reason to create a Finite State Machine was to have control over what move to make in a given situation. There was an initial attempt about having a LookerUp Table, which would give more control, to be able to pick which move to play. But with a LookerUp Table, there would need to be a lot of different combinations in order to obtain all possibilities. Also, since a Finite State Machine can be looked at as a type of LookerUp Table, and is a lot easier to code; a Finite State Machine strategy was chosen.

The most important aspects of the Finite State Machine we wanted to include were Memory One moves as well as Cooperation Rates. Memory One moves \cite{press_dyson_2012} are some of the best indicators of which move to play in the Prisoner's Dilemma. By taking into account the Memory One Strategies, $[C, C]$, $[C, D]$, $[D, C]$, $[D, D]$, a player can control what score their opponent can achieve. We used these 4 moves as a baseline for the Finite State Machine we wanted to create. 

After settling on the Four Memory One moves, the next important step was to look at Cooperation Rates from those moves. The Cooperation Rates for top players in the Tournament were similar to each other, so we simply decided to use that. According to \cite{harper_knight_jones_koutsovoulos_glynatsi_campbell_2017}, the top 80 players will always Cooperate with each other. There needed to be a way to ensure that our strategy would Always Cooperate with the other top strategies so we looked at the Cooperation Rates of those top 80 players that are currently in the Axelrod Library. This led to our initial design and subsequent strategies.

\begin{figure}

\tikzset{
->, 
node distance=3cm, 
every state/.style={circle, fill=gray!10, draw=green!60, radius=1cm}, 
initial text=$C$, 
}

\begin{tikzpicture}

    \node[state, initial] (p1) {1};
    \node[state] at (5, -3) (p2) {2};
    \node[state] at (7, -6) (p3) {3};
    \node[state] at (5, -9) (p4) {4};
    \node[state] at (0, -12) (p5) {5};
    \node[state] at (-5, -9) (p6) {6};
    \node[state] at (-7, -6) (p7) {7};
    \node[state] at (-5, -3) (p8) {8};
    
    \draw (p1) edge[bend right, above, color=red] node{C/C} (p2);
    \draw (p1) edge[bend right, above, color=blue] node{D/C} (p3);
    
    \draw (p2) edge[bend right, above, color=red] node{C/C} (p1);
    \draw (p2) edge[bend left, above, color=blue] node{D/C} (p3);
    
    \draw (p3) edge[bend left, above, color=red] node{C/C} (p2);
    \draw (p3) edge[bend left, above, color=blue] node{D/D} (p4);
    
    \draw (p4) edge[bend left, above, color=red] node{C/C} (p5);
    \draw (p4) edge[bend left, above, color=blue] node{D/D} (p6);
    
    \draw (p5) edge[bend right, above, color=red] node{C/C} (p2);
    \draw (p5) edge[bend right, above, color=blue] node{D/D} (p3);
    
    \draw (p6) edge[bend right, above, color=red] node{C/C} (p2);
    \draw (p6) edge[bend left, above, color=blue] node{D/D} (p7);
    
    \draw (p7) edge[bend left, above, color=red] node{C/D} (p4);
    \draw (p7) edge[bend right, above, color=blue] node{D/D} (p8);
    
    \draw (p8) edge[bend left, above, color=red] node{C/C} (p4);
    \draw (p8) edge[loop above, color=blue] node{D/D} (p4);
    
\end{tikzpicture}

\caption{FirstPrac Intro Diagram}
\label{fig:1}       
\end{figure}
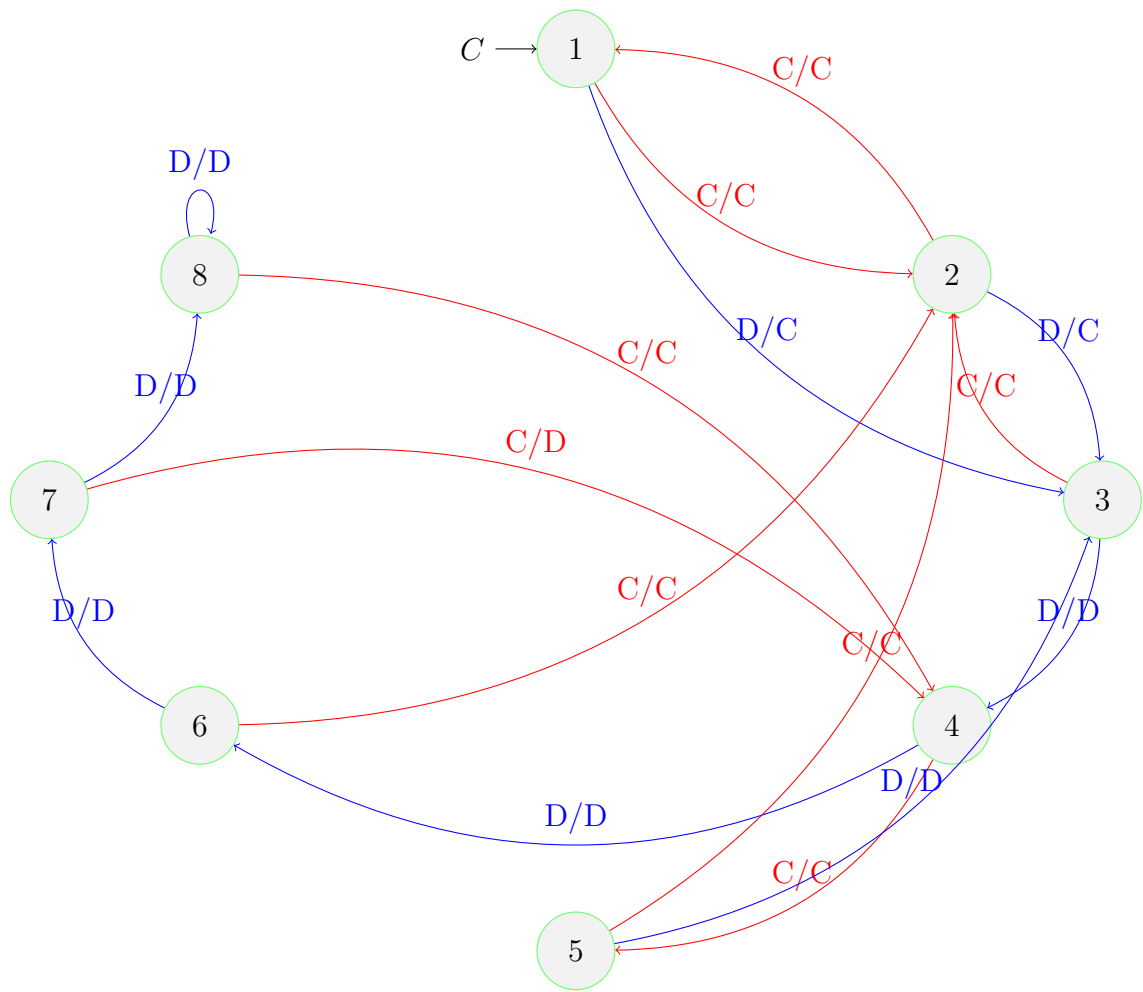

\section{Design}
\subsection{First Attempt}
The first attempt at making a Finite State Machine was to make one with a $100\%$ Cooperation Rate after a $[C, C]$ move and a $50\%$ Cooperation Rate for the other three moves, $[C, D]$, $[D, C]$, and $[D, D]$. The initial thought process for this strategy was to always Cooperate after a $[C, C]$ move, which led to a type of loop from State 1 to State 2 back to State 1. Then from the other states, there was always an attempt to alternate between Cooperating and Defecting after $[C, D]$, $[D, C]$, and $[D, D]$ moves. The reason to create an FSM with these specified Cooperation Rates  was to get a sense of how the Cooperation Rates based on the Memory One Moves can affect a strategy's score in the Axelrod Tournament.

This led to an eight-state FSM called \emph{FirstPrac}. Figure 1 shows a modified Finite State Diagram for the first attempt in the Axelrod Tournament. The eight states were created in order to try to maintain $50\%$ Cooperation rates after the other Memory One Moves. If there was a Cooperation after $[C, D]$, the next time there was a $[C, D]$, there would be a Defect move played. This led to the eight state FSM.

Figure 1 is my modified Finite State Machine diagram to write out the different states. Writing the diagram in this formed help me visualize what moves to play and how each state and move were connected to one another. The first move was to initially Cooperate and move to State 1. Then from State 1, if the opponent Cooperated, the player would Cooperate and move to State 2. From State 1, if the opponent Defected, the player would Cooperate and move to State 3. The rest of the diagram follows this system.

\begin{figure}

\tikzset{
->, 
node distance=3cm, 
every state/.style={circle, fill=gray!10, draw=green!60, radius=1cm}, 
initial text=$C$, 
}

\begin{tikzpicture}

    \node[state, initial] (p1) {1};
    \node[state] at (5, -3) (p2) {2};
    \node[state] at (7, -6) (p3) {3};
    \node[state] at (5, -9) (p4) {4};
    \node[state] at (0, -12) (p5) {5};
    \node[state] at (-5, -9) (p6) {6};
    \node[state] at (-7, -6) (p7) {7};
    \node[state] at (-5, -3) (p8) {8};
    
    \draw (p1) edge[bend right, above, color=red] node{C/C} (p2);
    \draw (p1) edge[bend right, above, color=blue] node{D/C} (p3);
    
    \draw (p2) edge[bend right, above, color=red] node{C/C} (p1);
    \draw (p2) edge[bend left, above, color=blue] node{D/C} (p3);
    
    \draw (p3) edge[bend left, above, color=red] node{C/C} (p2);
    \draw (p3) edge[bend left, above, color=blue] node{D/D} (p4);
    
    \draw (p4) edge[bend left, above, color=red] node{C/C} (p5);
    \draw (p4) edge[bend left, above, color=blue] node{D/D} (p6);
    
    \draw (p5) edge[bend right, above, color=red] node{C/C} (p2);
    \draw (p5) edge[bend right, above, color=blue] node{D/D} (p3);
    
    \draw (p6) edge[bend right, above, color=red] node{C/C} (p2);
    \draw (p6) edge[bend left, above, color=blue] node{D/D} (p7);
    
    \draw (p7) edge[bend left, above, color=red] node{C/D} (p4);
    \draw (p7) edge[bend right, above, color=blue] node{D/D} (p8);
    
    \draw (p8) edge[bend left, above, color=red] node{C/C} (p4);
    \draw (p8) edge[loop above, color=blue] node{D/D} (p4);
    
\end{tikzpicture}

\caption{FirstPrac Diagram}
\label{fig:2}       
\end{figure}
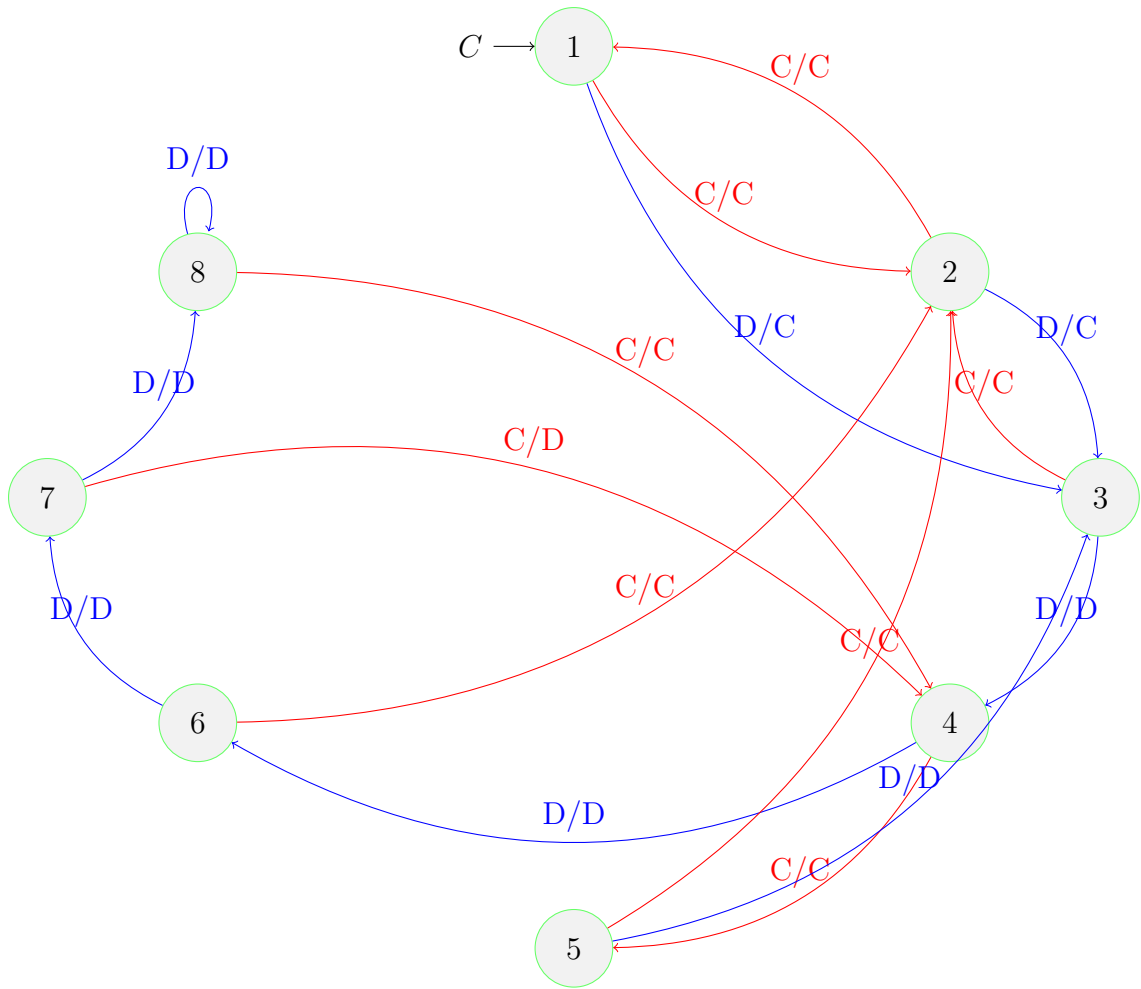

When running the \emph{FirstPrac} strategy against the short run time strategies, it finished in $107^{th}$ place and had Cooperation Rates of $100\%$, $73.6\%$, $100\%$, and $0\%$. This looks like that the strategy stayed in the same few states and was never in a scenario where it would Cooperate after a $[D, D]$ move. Figure 2 is another FSM diagram for \emph{FirstPrac}, this diagram is a more common diagram. It is a diagram that is used to describe the Finite State Machines in \cite{harper_knight_jones_koutsovoulos_glynatsi_campbell_2017}.

\subsection{Second Attempt}

The second attempt to create an FSM strategy used the Cooperation Rates from the top 80 players. According to \cite{harper_knight_jones_koutsovoulos_glynatsi_campbell_2017}, the top players always Cooperate with each other, so this new strategy had to be able to Cooperate with the other top players. The way to make sure our strategy always Cooperated with the top players was to take the average Cooperation Rates from the top 80 players. After running an Axelrod Tournament, a csv file with the Cooperation Rates after $[C, C]$, $[C, D]$, $[D, C]$, and $[D, D]$ moves are given. We used this information as a baseline for what we wanted our Cooperation Rates to look like. The Cooperation Rates we obtained from the Top 80 players, were $97.5\%$ after a $[C, C]$ move, $40.3\%$ after a $[C, D]$ move, $49.3\%$ after a $[D, C]$ move, and $17.6\%$ after a $[D, D]$ move. With this information, we rounded those percentages so we looked to create an FSM, with Cooperation Rates of $100\%$, $40\%$, $50\%$, and $20\%$. This led to the creation of the second strategy, a ten-state FSM, called \emph{SecondPrac}. Since there were more complex percentages in \emph{SecondPrac} than there was in \emph{FirstPrac}, there needed to be more states in order to try to maintain the Cooperation Rates from the top 80 players.

\begin{table}[ht]
\caption{FirstPrac} 
\begin{filecontents*}{csvs/thesis_1st_prac_2.csv}
Rank,Name,Median_score,Cooperation_rating,Wins,Initial_C_rate,CC_rate,CD_rate,DC_rate,DD_rate,CC_to_C_rate,CD_to_C_rate,DC_to_C_rate,DD_to_C_rate
100,Soft Go By Majority,2.582614679,0.82596789,16,1,0.743426606,0.082541284,0.043958716,0.130073394,1,0.879264822,0.741463658,0
101,Punisher,2.581892202,0.654369266,69,1,0.637672018,0.016697248,0.081493119,0.264137615,1,0,0.24639031,0
102,Win-Stay Lose-Shift: C,2.579633028,0.771711009,22,1,0.645561927,0.126149083,0.103408257,0.124880734,1,0,0,1
103,Nice Average Copier,2.579174312,0.784600917,29,1,0.712034404,0.072566514,0.057266055,0.158133028,0.882050381,0.528335037,0.591165272,0.427828402
104,Hard Tit For 2 Tats,2.572889908,0.839165138,0,1,0.771944954,0.067220183,0.024208716,0.136626147,1,0.705654542,0.385614881,0.300714558
105,FirstPrac,2.568555046,0.860983945,0,1,0.784791284,0.076192661,0.019465596,0.119550459,1,0.737543593,1,0
106,Tit For 2 Tats,2.565986239,0.866389908,0,1,0.78740367,0.078986239,0.017543578,0.116066514,1,0.76657814,1,0
107,N Tit(s) For M Tat(s): 3; 2,2.565940367,0.805142202,24,1,0.74708945,0.058052752,0.033204128,0.16165367,1,0.738320275,0.275340314,0.166485944
108,Hard Tit For Tat,2.564220183,0.643059633,72,1,0.628344037,0.014715596,0.080493119,0.276447248,1,0,0.245472817,0
109,Tricky Level Punisher,2.560424312,0.809889908,25.5,1,0.71490367,0.094986239,0.05646789,0.133642202,1,0.893105106,0.039151358,0.021200975
110,Thumper,2.556020642,0.701704128,54.5,1,0.667791284,0.033912844,0.063139908,0.235155963,1,0,0.574250552,0
111,Inverse,2.544736239,0.683419725,70,1,0.652697248,0.030722477,0.068050459,0.248529817,0.953698956,0,0.574012444,0
112,Two Tits For Tat,2.543681193,0.665442661,70,1,0.641004587,0.024438073,0.071495413,0.263061927,1,0,0.376726308,0
113,Eventual Cycle Hunter,2.523727064,0.905688073,15,1,0.758616972,0.147071101,0.037802752,0.056509174,0.997237817,0.983310181,0.112698722,0.134350246
114,Second by Tranquilizer,2.513474771,0.679637615,81.5,1,0.590442661,0.089194954,0.106167431,0.214194954,0.892670503,0.569050985,0.948300529,0.436495309
115,AdaptorBrief,2.4925,0.60356422,107.5,0.986697248,0.550614679,0.052949541,0.113126147,0.283309633,0.971515081,0.344596793,0.536660633,0.107945199
116,First by Nydegger,2.481376147,0.951949541,14,1,0.762899083,0.189050459,0.035827982,0.012222477,0.983902373,0.957331108,0.79735801,0.70479064
117,Random Hunter,2.415114679,0.946261468,12,1,0.757050459,0.189211009,0.02318578,0.030552752,0.994816627,0.982777822,0.486960698,0.134616838
118,First by Graaskamp: 0.05,2.402763761,0.702970183,99.5,1,0.541587156,0.161383028,0.121380734,0.175649083,0.929608398,0.781079618,0.85900449,0.34365179
119,Raider,2.390860092,0.430243119,89,0,0.378018349,0.052224771,0.172126147,0.397630734,0.991437039,0.467812262,0.663107961,0.321331064
120,Cycle Hunter,2.390573394,0.927146789,11,1,0.737261468,0.189885321,0.026516055,0.046337156,0.992380375,0.973703517,0,0
121,UsuallyCooperates,2.387236239,0.956704128,0,1,0.74171789,0.214986239,0.028956422,0.01433945,0.915949011,1,1,1
122,Sneaky Tit For Tat,2.384575688,0.76762844,51,1,0.581880734,0.185747706,0.102206422,0.130165138,0.900530132,0.578759217,0.734110803,0.682502815
123,Defector Hunter,2.38016055,0.988724771,0,1,0.78887156,0.199853211,0.000211009,0.01106422,1,0.98538511,1,0
124,Second by Gladstein,2.37891055,0.577740826,54.5,0,0.494885321,0.082855505,0.118869266,0.303389908,0.874029352,0.439351515,1,0.36661827
125,Second by Tester,2.376376147,0.576727064,56,0,0.493986239,0.082740826,0.119309633,0.303963303,0.876876702,0.434968641,1,0.365667072
126,AdaptorLong,2.373451835,0.447048165,102,0.994036697,0.337747706,0.109300459,0.203155963,0.349795872,0.93812356,0.01644159,0.012006376,0.319591307
127,Hard Prober,2.363933486,0.412807339,83.5,0,0.354594037,0.058213303,0.177396789,0.409795872,0.887001997,0.45362807,0.513163132,0.37069044
128,Ripoff,2.361548165,0.567672018,63.5,0,0.482392202,0.085279817,0.119928899,0.312399083,0.972904054,0.454338703,1,0.31749151
129,Prober 2,2.361307339,0.722853211,55.5,0,0.613502294,0.109350917,0.060486239,0.21666055,1,0.63928607,1,0.436637058
130,Alternator Hunter,2.353291284,0.977752294,3,1,0.765279817,0.212472477,0.008912844,0.013334862,0.994603175,0.994968553,0,0
131,Second by Colbert,2.345263761,0.81053211,63,1,0.640674312,0.169857798,0.058715596,0.130752294,0.971720668,0.558353225,0.781997394,0.603536635
132,Cooperator,2.339793578,1,0,1,0.779103211,0.220896789,0,0,1,1,0,0
133,Average Copier,2.313830275,0.644770642,63.5,0.503211009,0.547451835,0.097318807,0.078899083,0.276330275,0.857227663,0.458889052,0.671424798,0.334069057
134,First by Downing,2.271376147,0.294759174,113.5,0,0.257864679,0.036894495,0.19756422,0.507676606,0.962296768,0.607792518,0.241193786,0.303156801
135,DoubleResurrection,2.263038991,0.516981651,102.5,1,0.319649083,0.197332569,0.205172018,0.27784633,0.841133841,0,1,0.171517563
136,Willing,2.255493119,0.957041284,21,0.505504587,0.736011468,0.221029817,0.002272936,0.04068578,1,1,1,0
137,Adaptive,2.251674312,0.522873853,106.5,1,0.360727064,0.162146789,0.173493119,0.303633028,0.927697991,0.942537147,0.125949814,0.205985151
138,EasyGo,2.227133028,0.919550459,43,0,0.614667431,0.304883028,0.075777523,0.004672018,1,1,0,1
139,Suspicious Tit For Tat,2.221261468,0.552415138,107,0,0.438513761,0.113901376,0.116364679,0.331220183,1,0,1,0
140,Risky QLearner,2.219392202,0.862722477,63,0.953669725,0.644497706,0.218224771,0.036777523,0.1005,0.947534087,0.929186759,0.951551205,0.63532378
141,Cautious QLearner,2.217740826,0.861052752,61.5,0.953669725,0.645337156,0.215715596,0.036110092,0.102837156,0.949851434,0.928557958,0.95000528,0.630650037
142,Arrogant QLearner,2.216708716,0.86059633,61,0.946330275,0.645330275,0.215266055,0.036619266,0.102784404,0.949108592,0.928855401,0.950410292,0.636164334
143,Hesitant QLearner,2.211639908,0.862855505,62,0.946788991,0.643224771,0.219630734,0.036454128,0.100690367,0.948332074,0.927027476,0.950922225,0.637293416
144,Prober,2.209449541,0.38396789,78.5,0,0.313954128,0.070013761,0.162208716,0.453823394,0.835121036,0.445208834,0.70625675,0.268053658
145,Remorseful Prober: 0.1,2.207522936,0.47134633,141,1,0.351357798,0.119988532,0.15518578,0.37346789,0.898152325,0,0.901546186,0.311869712
146,Detective,2.201272936,0.429087156,65,1,0.358321101,0.070766055,0.138669725,0.432243119,0.763791332,0.413903013,0.719079894,0.259176623
147,Opposite Grudger,2.200676606,0.924495413,39,0,0.700575688,0.223919725,0.00465367,0.070850917,1,1,1,0
148,Random Tit for Tat: 0.5,2.196559633,0.523513761,89,1,0.331084862,0.192428899,0.180779817,0.295706422,0.737716166,0.238404442,0.859161981,0.294111396
149,Worse and Worse 2,2.185516055,0.311016055,121.5,1,0.17991055,0.131105505,0.238277523,0.450706422,0.689363923,0.284422607,0.335090416,0.189846324
150,ZD-Mem2,2.18375,0.503181193,117.5,1,0.383330275,0.119850917,0.134518349,0.362300459,0.875696238,0.164284327,0.790724998,0.120132648
151,"First by Feld: 1.0, 0.5, 200",2.178669725,0.369334862,171.5,1,0.300988532,0.06834633,0.161350917,0.46931422,0.872478765,0,0.792226906,0
152,Worse and Worse,2.171502294,0.899688073,74,0.998165138,0.6055,0.294188073,0.063091743,0.037220183,0.915124077,0.890795626,0.872632239,0.851758009
153,Prober 4,2.166525229,0.226786697,127,1,0.186139908,0.040646789,0.208277523,0.56493578,0.562690152,0.321109462,0.439893235,0.161609306
154,Prober 3,2.161387615,0.258245413,131,0,0.184268349,0.073977064,0.216947248,0.524807339,0.969670158,0,0.624050399,0.178933548
155,Stochastic Cooperator,2.144633028,0.561977064,96,1,0.381481651,0.180495413,0.140956422,0.297066514,0.93431779,0.232005503,0.25844816,0.418829024
156,Knowledgeable Worse and Worse,2.142006881,0.496974771,98,0.994036697,0.282892202,0.214082569,0.197486239,0.305538991,0.739521177,0.565574431,0.479888262,0.301368337
157,Hard Go By Majority: 5,2.141639908,0.51412156,105.5,0,0.423699541,0.090422018,0.095667431,0.390211009,1,0.655503655,0.802194572,0
158,Hard Go By Majority,2.136708716,0.501360092,104,0,0.406321101,0.095038991,0.104646789,0.393993119,1,0.792195459,0.786283181,0
159,First by Tullock,2.130172018,0.44937156,151,1,0.376850917,0.072520642,0.113511468,0.437116972,0.819679314,0.58505349,0.621220646,0.190097365
160,ALLCorALLD,2.128383028,0.608256881,81.5,0.608256881,0.472509174,0.135747706,0.077745413,0.313997706,1,1,0,0
161,Pun1,2.123623853,0.773415138,55,0,0.519949541,0.253465596,0.084162844,0.142422018,1,0.354586404,1,1
162,Fortress3,2.107672018,0.321795872,98,0,0.135094037,0.186701835,0.256325688,0.42187844,1,0,0,0.462756552
163,Fortress4,2.096961009,0.22571789,105,0,0.083110092,0.142607798,0.268295872,0.505986239,1,0,0,0.302047435
164,Calculator,2.081926606,0.313669725,147.5,1,0.246158257,0.067511468,0.163105505,0.523224771,0.929032098,0,0.613029586,0
165,First by Joss: 0.9,2.08108945,0.414928899,171,1,0.336672018,0.078256881,0.121105505,0.463965596,0.898143528,0,0.89242491,0
166,Meta Hunter Aggressive: 7 players,2.079415138,0.173302752,117,1,0.093958716,0.079344037,0.242607798,0.58408945,0.742459018,0.741827037,0.151503828,0.074619545
167,Naive Prober: 0.1,2.078600917,0.414451835,171,1,0.335942661,0.078509174,0.121584862,0.463963303,0.900106867,0,0.899168457,0
168,Cooperator Hunter,2.071834862,0.955334862,43,1,0.620956422,0.33437844,0.041408257,0.003256881,0.961709672,1,0,1
169,Hard Go By Majority: 40,2.071318807,0.458768349,124,0,0.401463303,0.057305046,0.08162844,0.459603211,1,0.7756982,0.745918326,0
170,Hard Go By Majority: 20,2.057763761,0.447050459,127,0,0.39737844,0.049672018,0.078587156,0.474362385,1,0.735747006,0.732388685,0
171,Hard Go By Majority: 10,2.045332569,0.430483945,133.5,0,0.385690367,0.044793578,0.079399083,0.490116972,1,0.691379503,0.699728462,0
172,Cycler CCCDCD,2.04266055,0.67,72.5,1,0.321357798,0.348642202,0.187002294,0.142997706,0.517724271,0.544733453,1,1
173,SolutionB1,2.04233945,0.917832569,41,0,0.64446789,0.273364679,0.006658257,0.075509174,1,1,0.451593225,0.909321624
174,Stochastic WSLS: 0.05,2.03815367,0.550928899,80.5,1,0.286793578,0.264135321,0.181580275,0.267490826,0.946482717,0.053503099,0.051734432,0.950189228
175,Winner21,2.028899083,0.347165138,175,0,0.315573394,0.031591743,0.107321101,0.545513761,1,0,0.554991962,0.2322756
176,Cycler CCCCCD,2.027236239,0.835,65.5,1,0.478857798,0.356142202,0.105899083,0.059100917,0.784009025,0.867365357,1,1
177,Cycler CCCD,2.025194954,0.75,77,1,0.390151376,0.359848624,0.150699541,0.099300459,0.625049778,0.763354599,1,1
178,SolutionB5,2.01587156,0.38784633,157.5,0,0.32833945,0.059506881,0.104525229,0.50762844,1,0,0.784733495,0
179,TF2,2.012626147,0.19162844,174,1,0.160275229,0.031353211,0.180527523,0.627844037,0.518371576,0,0.572191775,0.081315776
180,"ZD-SET-2: 0.25, 0.0, 2",2.009816514,0.41421789,99,1,0.231722477,0.182495413,0.181208716,0.404573394,0.744177779,0.253401067,0.501135608,0.255722105
181,Cycler CCD,1.997155963,0.67,63.5,1,0.304513761,0.365486239,0.187711009,0.142288991,0.409139545,0.637450988,1,1
182,$\pi$,1.980768349,0.80381422,73,1,0.347323394,0.456490826,0.184901376,0.011284404,0.588862042,1,0.882996748,1
183,SelfSteem,1.975401376,0.472075688,78.5,0.490366972,0.258236239,0.21383945,0.168091743,0.359832569,0.716862362,0.393533623,0.594274877,0.275691625
184,AntiCycler,1.973417431,0.9,61.5,1,0.538830275,0.361169725,0.064543578,0.035456422,0.848260583,0.86985421,0.90238509,0.987669065
185,Meta Winner Memory One: 36 players,1.968830275,0.016711009,174.5,1,0.011344037,0.005366972,0.237545872,0.745743119,0.519369164,0.411897208,0.053782949,0.017553225
186,$e$,1.955768349,0.785447248,77.5,1,0.327142202,0.458305046,0.190770642,0.02378211,0.57065079,1,0.760181914,1
187,ThueMorseInverse,1.955068807,0.5,87,1,0.176683486,0.323316514,0.23096789,0.26903211,0.291145867,0.395047446,0.586417336,0.755721788
188,Alternator,1.954587156,0.5,84.5,1,0.158291284,0.341708716,0.245240826,0.254759174,0,0,1,1
189,Tricky Cooperator,1.954438073,0.827665138,73,1,0.406247706,0.421417431,0.14134633,0.030988532,0.886975508,1,0,1
190,Desperate,1.923302752,0.361268349,88.5,0.501376147,0.054727064,0.306541284,0.278077982,0.36065367,0,0,0,1
191,"ZD-Extort-2: 0.1111111111111111, 0.5",1.922270642,0.266424312,184.5,1,0.215614679,0.050809633,0.136004587,0.597571101,0.887009651,0.512066238,0.327517629,0
192,Bully,1.92065367,0.587545872,83,0,0.134908257,0.452637615,0.274557339,0.137896789,0,1,0,1
193,Better and Better,1.917006881,0.100616972,117.5,0.000917431,0.026566514,0.074050459,0.234215596,0.665167431,0.148002297,0.133750206,0.100853174,0.096075476
194,ThueMorse,1.915034404,0.5,71.5,0,0.171850917,0.328149083,0.224431193,0.275568807,0.299575281,0.398655497,0.648840816,0.747982407
195,CollectiveStrategy,1.914919725,0.014793578,200,1,0.013582569,0.001211009,0.223520642,0.76168578,0.589189428,0,0,0.224939683
196,Predator,1.914025229,0.088584862,167,1,0.071733945,0.016850917,0.19715367,0.714261468,0.744083138,0,0.236448142,0.040041898
197,Cycler DC,1.912694954,0.5,67,0,0.144571101,0.355428899,0.244855505,0.255144495,0,0,1,1
198,Win-Shift Lose-Stay: D,1.912522936,0.542119266,89.5,0,0.206775229,0.335344037,0.209357798,0.248522936,0,1,1,0
199,First by Anonymous,1.912293578,0.500568807,78,0.496330275,0.20115367,0.299415138,0.20125,0.298181193,0.499351475,0.506075863,0.499847415,0.500980274
200,"ZD-Extort-2 v2: 0.125, 0.5, 1",1.909036697,0.261238532,182.5,1,0.211130734,0.050107798,0.135809633,0.602951835,0.872095637,0.443841375,0.374942128,0
201,"ZD-Mischief: 0.1, 0.0, 1",1.895298165,0.097733945,190,1,0.075575688,0.022158257,0.190958716,0.711307339,0.789265611,0.59673443,0.109058598,0
202,Cycler DDC,1.894266055,0.33,106.5,0,0.08062844,0.24937156,0.246137615,0.423862385,0,0,0.426459551,0.623882767
203,Random: 0.5,1.891479358,0.501800459,77.5,0.497706422,0.201752294,0.300048165,0.199669725,0.298529817,0.496555699,0.501706161,0.503646074,0.511139753
204,UsuallyDefects,1.889323394,0.07268578,151.5,0,0.024894495,0.047791284,0.222002294,0.705311927,0,0,0,0.263059195
205,TF1,1.888509174,0.095736239,173,1,0.068119266,0.027616972,0.195993119,0.708270642,0.431093401,0.625715541,0.38962152,0.081840282
206,Anti Tit For Tat,1.887305046,0.597048165,73.5,1,0.14216055,0.454887615,0.262706422,0.140245413,0,1,0,1
207,Tricky Defector,1.885997706,0.391502294,102,0,0.061662844,0.32983945,0.273245413,0.335252294,0,1,0,0.299269199
208,$\phi$,1.885699541,0.702376147,67,1,0.233350917,0.469025229,0.222116972,0.075506881,0.451400698,1,0.325393642,1
209,"ZD-Extort-4: 0.23529411764705882, 0.25, 1",1.883543578,0.161261468,189.5,1,0.118387615,0.042873853,0.170245413,0.668493119,0.637684996,0,0.468859415,0
210,Negation,1.881318807,0.594355505,78,0.513761468,0.138869266,0.455486239,0.266334862,0.139309633,0,1,0,1
211,"ZD-Extort3: 0.11538461538461539, 0.3333333333333333, 1",1.880458716,0.199763761,187.5,1,0.157837156,0.041926606,0.151119266,0.649116972,0.845446973,0.50710785,0.271534327,0
212,Aggravater,1.877637615,0.077555046,202.5,0,0.076724771,0.000830275,0.18059633,0.741848624,1,0,0.132773591,0
213,"ZD-Extortion: 0.2, 0.1, 1",1.876158257,0.116451835,195,1,0.086318807,0.030133028,0.183850917,0.699697248,0.630090909,0.208482809,0.283343976,0
214,"Gradual Killer: (D, D, D, D, D, C, C)",1.875412844,0.582360092,78,0,0.427727064,0.154633028,0.042561927,0.375077982,1,0.791286251,0.162116277,0.143580508
215,Hopeless,1.867798165,0.743286697,69,0.494954128,0.255298165,0.487988532,0.211075688,0.045637615,0,1,1,1
216,"Bush Mosteller: 0.5, 0.5, 3.0, 0.5",1.858566514,0.295412844,107,0.486697248,0.132394495,0.163018349,0.189387615,0.515199541,0.536312083,0.284400953,0.290862421,0.267156053
217,Defector,1.817201835,0,202,0,0,0,0.205259174,0.794740826,0,0,0,0
218,Handshake,1.815538991,0.074481651,190,1,0.041866972,0.032614679,0.19031422,0.735204128,0.917541799,0.992647524,0,0.432267953
\end{filecontents*}
\csvreader[
  longtable=llr,
  table head= 
    \toprule\bfseries Rank &\bfseries Name &\bfseries Median Score\\ \midrule\endhead
    \bottomrule\endfoot,
  late after line=\\,
  filter= \value{csvrow} < 15,
  before reading={\catcode`\#=12},
  after reading={\catcode`\#=6}
]{csvs/thesis_1st_prac_2.csv}{1=\Rank,2=\Name, 3=\Median_score}{\Rank & \Name & \Median_score}
\end{table}
{\addtocounter{table}{-1}} 

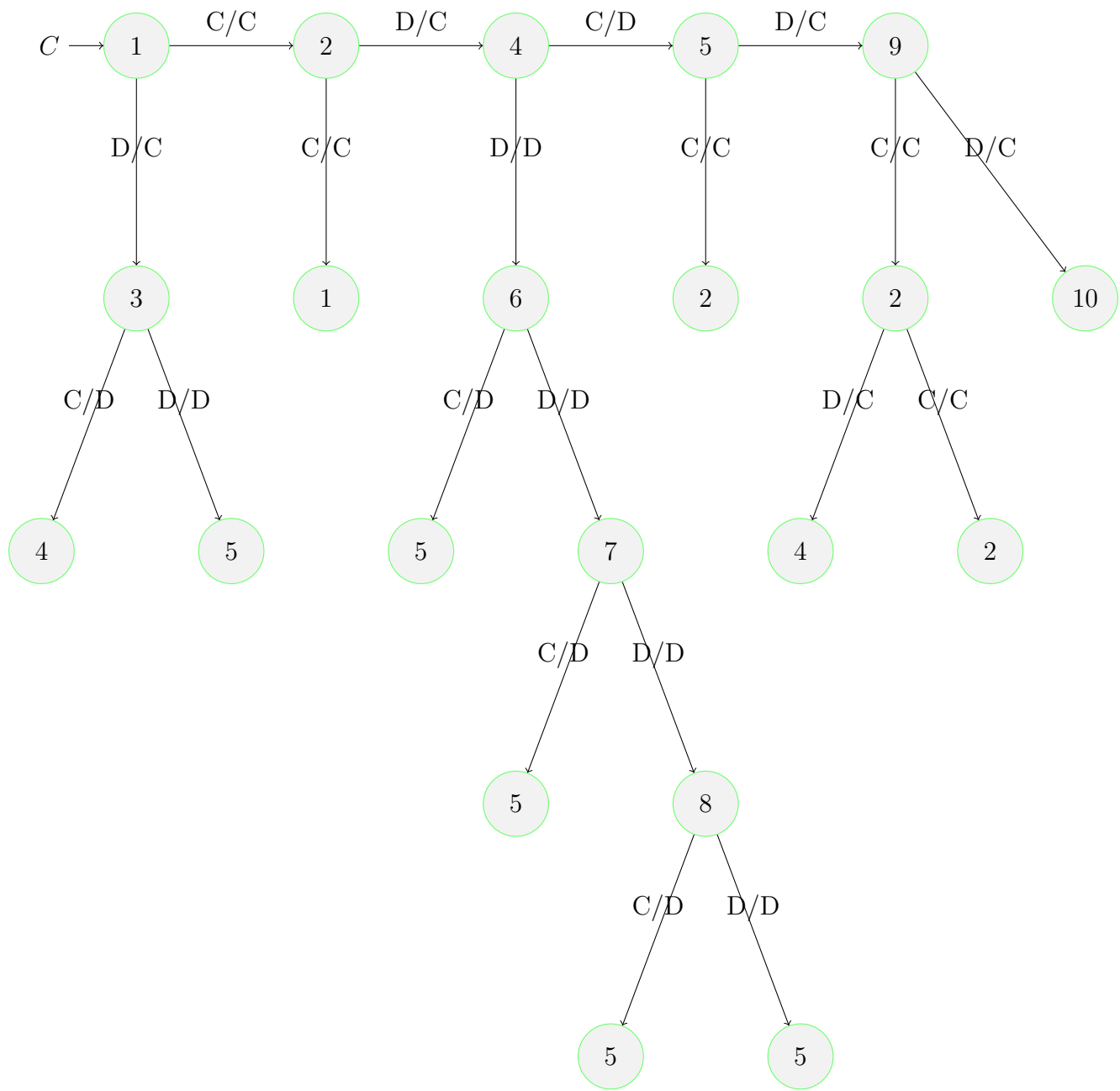
\begin{figure}
\tikzset{
->, 
node distance=3cm, 
every state/.style={circle, fill=gray!10, draw=green!60, radius=1cm}, 
initial text=$C$, 
}

\begin{tikzpicture}

    \node[state, initial] (p1) {1};
    \node[state] at (3, 0) (p2) {2};
    \node[state] at (6, 0) (p4) {4};
    \node[state] at (9, 0) (p5) {5};
    \node[state] at (12, 0) (p7) {9};
    \node[state] at (0, -4) (p9) {3};
    \node[state] at (-1.5, -8) (p10) {4};
    \node[state] at (1.5, -8) (p11) {5};
    \node[state] at (3, -4) (p3) {1};
    \node[state] at (6, -4) (p6) {6};
    \node[state] at (4.5, -8) (p12) {5};
    \node[state] at (7.5, -8) (p13) {7};
    \node[state] at (6, -12) (p14) {5};
    \node[state] at (9, -12) (p15) {8};
    \node[state] at (7.5, -16) (p16) {5};
    \node[state] at (10.5, -16) (p17) {5};
    \node[state] at (9, -4) (p8) {2};
    \node[state] at (12, -4) (p19) {2};
    \node[state] at (15, -4) (p18) {10};
    \node[state] at (13.5, -8) (p20) {2};
    \node[state] at (10.5, -8) (p21) {4};

    \draw (p1) edge[above] node{D/C} (p9);
    \draw (p1) edge[above] node{C/C} (p2);
    
    \draw (p9) edge[above] node{C/D} (p10);
    \draw (p9) edge[above] node{D/D} (p11);
    
    \draw (p2) edge[above] node{D/C} (p4);
    \draw (p2) edge[above] node{C/C} (p3);
    
    \draw (p4) edge[above] node{C/D} (p5);
    \draw (p4) edge[above] node{D/D} (p6);
    
    \draw (p6) edge[above] node{C/D} (p12);
    \draw (p6) edge[above] node{D/D} (p13);
    
    \draw (p13) edge[above] node{C/D} (p14);
    \draw (p13) edge[above] node{D/D} (p15);
    
    \draw (p15) edge[above] node{C/D} (p16);
    \draw (p15) edge[above] node{D/D} (p17);
    
    \draw (p5) edge[above] node{C/C} (p8);
    \draw (p5) edge[above] node{D/C} (p7);
    
    \draw (p7) edge[above] node{D/C} (p18);
    \draw (p7) edge[above] node{C/C} (p19);
    
    \draw (p19) edge[above] node{C/C} (p20);
    \draw (p19) edge[above] node{D/C} (p21);
    
\end{tikzpicture}
\caption{SecondPrac Intro Diagram}
\label{fig:3}       
\end{figure}

Another important idea from \cite{harper_knight_jones_koutsovoulos_glynatsi_campbell_2017} was that the top strategies always Cooperate during the first move. Since the top strategies always Cooperate during the first move, there must be a loop where they always Cooperate with the player that Cooperates with them. So the new strategy would Cooperate first, if the opponent Cooperated, the FSM would move to a second state State 2, and Cooperate. If the opponent cooperated again, the FSM would go back to the first state and Cooperate so there was a loop, where both players would Cooperate until there was a Defect. The idea was so that if there was a top player involved, those strategies would always Cooperate with each other, receiving a score of 3 all the time. Figure 3 shows my initial attempt at a Finite State Machine Diagram that followed the percentages from the top 80 players.

If the opponent initially Defected, the FSM would then go to state 3 and also Defect. So after a $[C, D]$ move, there was one instance of Cooperating and one instance of Defecting. 

That was important, to keep the number of Cooperates and Defects in order to try to maintain the correct percentages. From State 3, if the opponent Cooperated, there would be a $[D, C]$ move, which opened up a new state, state 4 where the FSM would Cooperate. This was an attempt to get the Cooperation rate after $[D, C]$ to be $50\%$. And if the opponent Defected on state 3, the FSM would move to a new state, State 5 and Defect. The majority of the time the FSM would Defect after a $[D, D]$ move because the percentage of the top scores Defecting was $20\%$.

From State 4, if the opponent Cooperated, the FSM would move to State 5 and Cooperate. And from state 5 if the opponent Cooperated, the FSM would go back to state 2 in order to start a new loop of Cooperation. If the opponent Defected in State 4, the FSM would move to a new state 6, where the next move was to Defect. 

State 6 is a state that the FSM had continued to Defect after $[D, D]$ moves in order to maintain a $20\%$ Cooperation rate. This would continue into State 7 and if the opponent Defected in State 7, the FSM would move to a new state, State 8, and also Defect. Once in State 8, after Defecting four times after a $[D, D]$ move, the FSM would Cooperate to get the $20\%$ Cooperation rate. After finally Cooperating after a $[D, D]$, the FSM would move to State 5. Also, from State 6, State 7, and State 8, if the opponent Cooperated instead of Defecting, the FSM would move to State 5 and would alternate between Cooperating and Defecting.

From state 5, if the opponent Defected, this would create a $[C, D]$ move where the FSM would move to a new state, State 9 and would Cooperate. The new state was created to try to maintain the percentages from the Top players. In state 9, if the opponent Cooperated, that resulted in a $[C, C]$, in which the FSM would always Cooperated, and the FSM would move back to State 2, to go back into the Cooperation loop. And if the opponent Defected, this would create a $[C, D]$ move and to maintain the percentages, the FSM would move to state 10 and would Cooperate. From state 10, if the opponent Cooperated, this would create a $[C, C]$ move so the FSM will Cooperate, and will go back to State 2. And if the opponent Defected in State 10, the FSM will move back to State 4 and Defect.

In total, there were 20 total moves scenarios. There were 5 instances of $[C, C]$, where every time the FSM Cooperated. There were 5 instances of $[C, D]$, where the FSM would Cooperate $\frac{2}{5}$ times and Defect $\frac{3}{5}$ times. There were 5 instances of $[D, C]$, where the FSM would Cooperate $\frac{3}{5}$ times and Defect $\frac{2}{5}$ times. And there were 5 instances of $[D, D]$, where the FSM would Defect $\frac{4}{5}$ times and Cooperate $\frac{1}{5}$ times.

This strategy finished $35^{th}$ in the short run time strategies in the Axelrod Library. It had Cooperation rates of $100\%$ after $[C, C]$, $40\%$ after a $[C, D]$, $83.7\%$ after $[D, C]$, and a $46\%$ after $[D, D]$. So it was getting closer to our desired Cooperation Rates. Figure 4 shows the Finite State Machine Diagram that corresponded to the \emph{SecondPrac} strategy.

\begin{table}[ht]
\caption{SecondPrac Scores}

\csvreader[
  longtable=llr,
  table head= 
    \toprule\bfseries Rank &\bfseries Name &\bfseries Median Score\\ \midrule\endhead
    \bottomrule\endfoot,
  late after line=\\,
  filter= \value{csvrow} < 15,
  before reading={\catcode`\#=12},
  after reading={\catcode`\#=6}
]{csvs/thesis_2nd_prac_2.csv}{1=\Rank,2=\Name, 3=\Median_score}{\Rank & \Name & \Median_score}
\end{table}
{\addtocounter{table}{-1}} 

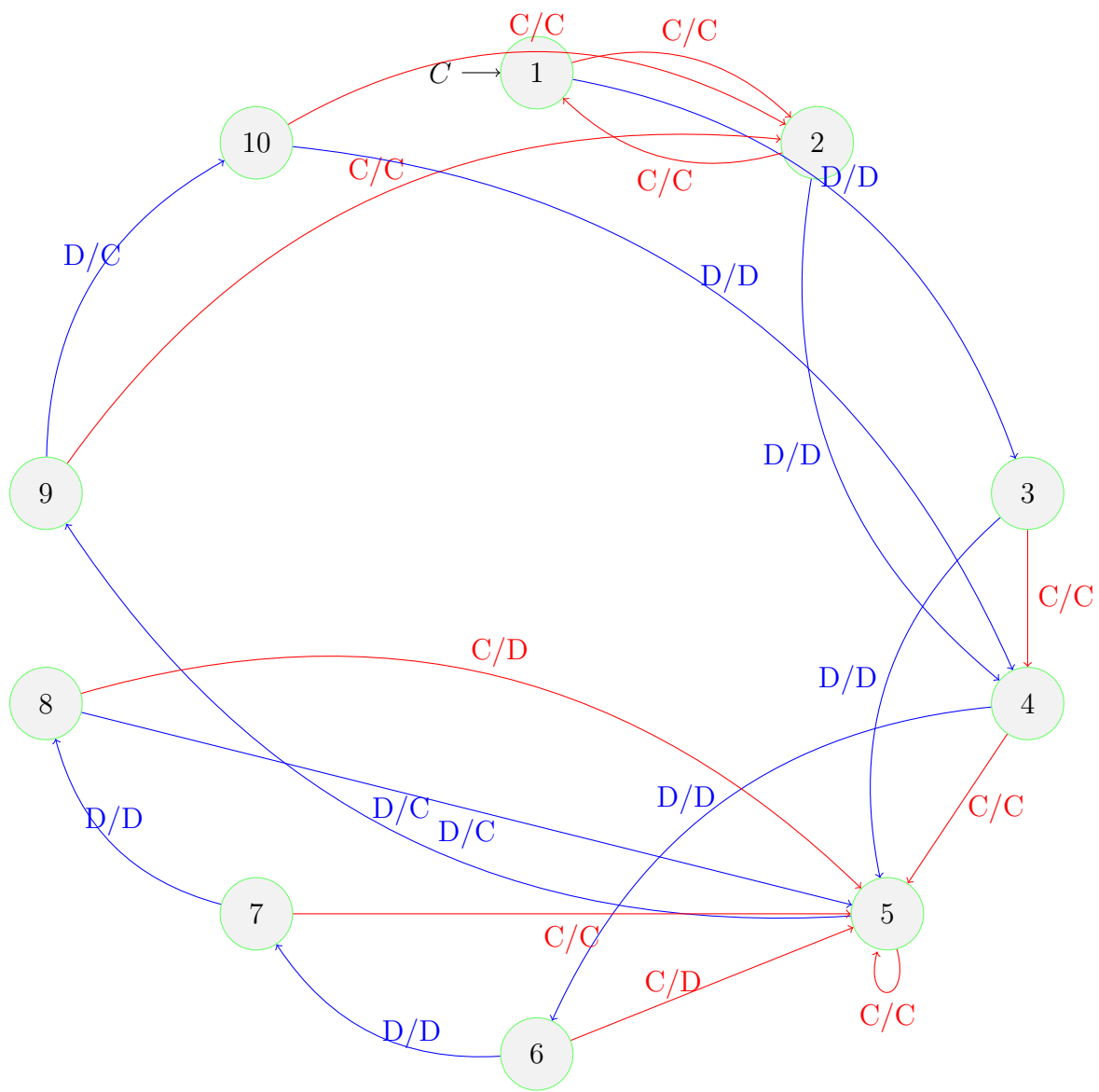
\begin{figure}

\tikzset{
->, 
node distance=3cm, 
every state/.style={circle, fill=gray!10, draw=green!60, radius=1cm}, 
initial text=$C$, 
}

\begin{tikzpicture}

    \node[state, initial] (p1) {$1$};
    \node[state] at (4, -1) (p2) {$2$};
    \node[state] at (7, -6) (p3) {$3$};
    \node[state] at (7, -9) (p4) {$4$};
    \node[state] at (5, -12) (p5) {$5$};
    \node[state] at (0, -14) (p6) {$6$};
    \node[state] at (-4, -12) (p7) {$7$};
    \node[state] at (-7, -9) (p8) {$8$};
    \node[state] at (-7, -6) (p9) {$9$};
    \node[state] at (-4, -1) (p10) {$10$};
    
    \draw (p1) edge[bend left, above, color=red] node{C/C} (p2);
    \draw (p1) edge[bend left, above, color=blue] node{D/D} (p3);
    
    \draw (p2) edge[bend left, below, color=red] node{C/C} (p1);
    \draw (p2) edge[bend right, left, color=blue] node{D/D} (p4);
    
    \draw (p3) edge[right, color=red] node{C/C} (p4);
    \draw (p3) edge[bend right, left, color=blue] node{D/D} (p5);
    
    \draw (p4) edge[right, color=red] node{C/C} (p5);
    \draw (p4) edge[bend right, left, color=blue] node{D/D} (p6);
    
    \draw (p5) edge[loop below, below, color=red] node{C/C} (p2);
    \draw (p5) edge[bend left, above, color=blue] node{D/C} (p9);
    
    \draw (p6) edge[left, color=red] node{C/D} (p5);
    \draw (p6) edge[bend left, right, color=blue] node{D/D} (p7);
    
    \draw (p7) edge[below, color=red] node{C/C} (p5);
    \draw (p7) edge[bend left, above, color=blue] node{D/D} (p8);
    
    \draw (p8) edge[bend left, above, color=red] node{C/D} (p5);
    \draw (p8) edge[below, color=blue] node{D/C} (p5);
    
    \draw (p9) edge[bend left, above, color=red] node{C/C} (p2);
    \draw (p9) edge[bend left, above, color=blue] node{D/C} (p10);
    
    \draw (p10) edge[bend left, above, color=red] node{C/C} (p2);
    \draw (p10) edge[bend left, above, color=blue] node{D/D} (p4);
    
\end{tikzpicture}
\caption{SecondPrac Diagram}
\label{fig:4}       
\end{figure}

\subsection{Other Attempts}
\subsubsection{SecondPrac Variants}
After seeing the success of this strategy, \emph{SecondPrac2} and \emph{SecondPrac3} were created after modifying a couple of moves from \emph{SecondPrac}. There was not much success with them. For \emph{SecondPrac2}, a move in State 9 was changed. Initially, if the opponent Defected from state 5, the FSM would move to State 9 and Cooperate; it was  changed to a Defect. This change altered the percentages as well as where \emph{SecondPrac2} scored. \emph{SecondPrac2} finished $57^{th}$ and had Cooperate Rates of $100\%$ after $[C, C]$, $0\%$ after $[C, D]$, $88.3\%$ after $[D, C]$, and $37.5\%$ after a $[D, D]$.

After modifying \emph{SecondPrac2} by switching a move in State 5. There was another modification, initially, if the opponent Cooperated in State 4, the FSM would move to State 5 and would Cooperate. After the change, the FSM would go to state 5, and would Defect. These two changes combined had a worse effect on the score. \emph{SecondPrac3} would finish $96^{th}$ and had Cooperation Rates of $98.4\%$ after a $[C, C]$, $0\%$ after a $[C, D]$, $57.0\%$ after a $[D, C]$, and $36.6\%$ after a $[D, D]$.

\subsection{ThirdPrac}

After failing to modify the \emph{SecondPrac} strategy that finished $34^{th}$ successfully, a new strategy, \emph{ThirdPrac} was created. \emph{ThirdPrac} was created to try to maintain the same percentages from the top $80$ strategies, but the states were completly altered. This resulted in an 8-state FSM. This FSM finished 58th and had Cooperation Rates of $100\%$ after a $[C, C]$ move, $44.1\%$ after a $[C, D]$ move, $91.9\%$ after a $[D, C]$ move, and $0\%$ after a $[D, D]$ move. It was at this point, to go back to \emph{SecondPrac} and evolve it.

\subsection{FourthPrac}

Initially when trying to evolve \emph{SecondPrac} using 500 generations, which is the default setting for the Axelrod Tournament, there would be an error that would occur around 40-60 generations on my computer. So I decided to evolve it using 50 generations, with a population size of 40, .1 mutation rate, a bottleneck of 10, 1 process, 10 repetitions, 20 turns, 0 noise, 4 Moran processes, and 10 states. This evolution worked and the \emph{FourthPrac} strategy was created. This was a 10-ten state FSM that finished in the top 5 of the Axelrod Library in the Short Run Time tournament. It had the following Cooperation Rates, $92.5\%$ after a $[C, C]$ move, $26.3\%$ after a $[C, D]$ move, $22.3\%$ after a $[D, C]$ move, and $37.1\%$ after a $[D, D]$ move.

\begin{table}[ht]
\caption{FourthPrac Scores}\label{tab:test}

\csvreader[
  longtable=llr,
  table head= 
    \toprule\bfseries Rank &\bfseries Name &\bfseries Median Score\\ \midrule\endhead
    \bottomrule\endfoot,
  late after line=\\,
  filter= \value{csvrow} < 15,
  before reading={\catcode`\#=12},
  after reading={\catcode`\#=6}
]{csvs/thesis_4th_prac.csv}{1=\Rank,2=\Name, 3=\Median_score}{\Rank & \Name & \Median_score}
\end{table}
{\addtocounter{table}{-1}}

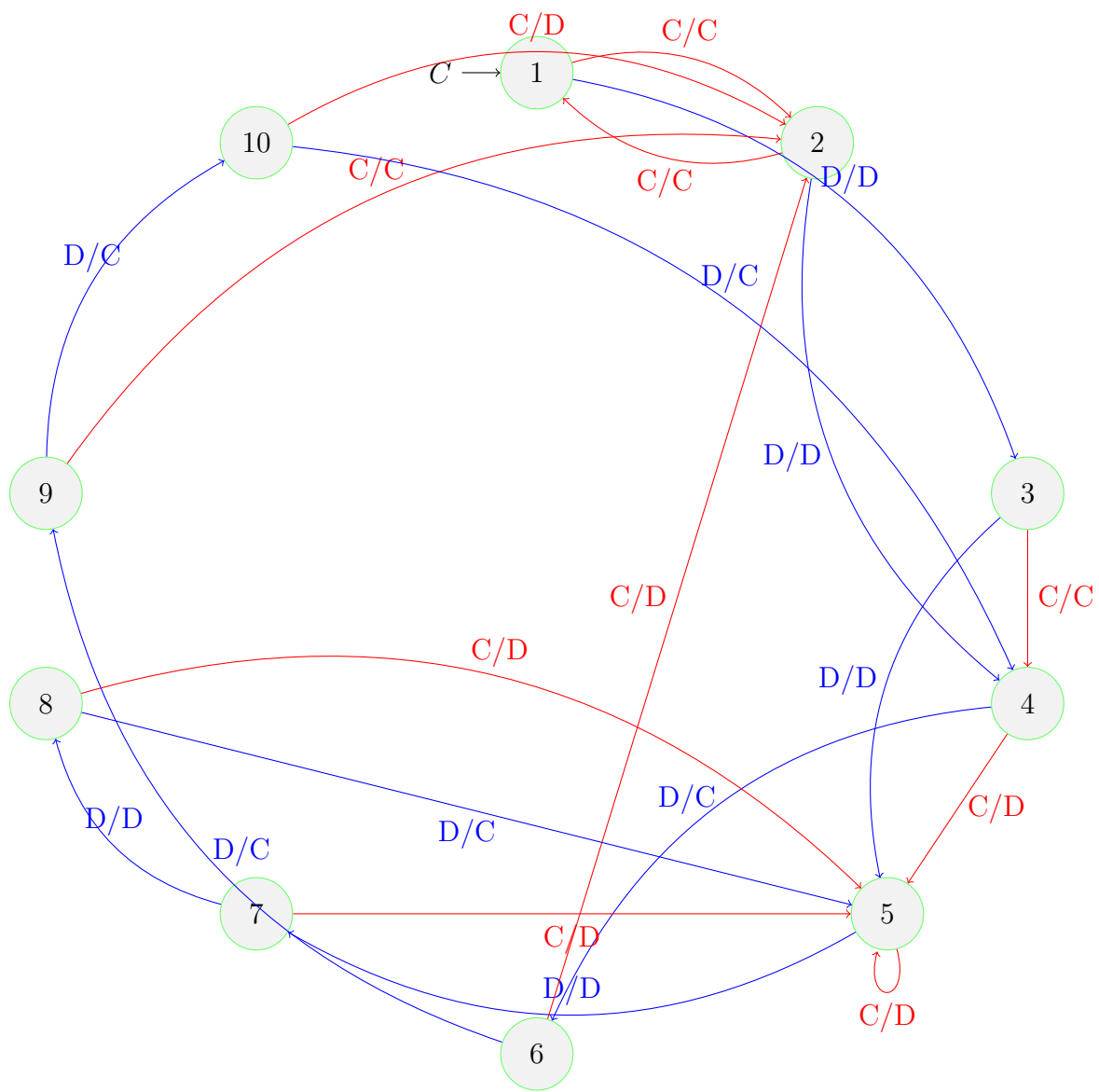
\begin{figure}

\tikzset{
->, 
node distance=1cm, 
every state/.style={circle, fill=gray!10, draw=green!60, radius=1cm}, 
initial text=$C$, 
}

\begin{tikzpicture}

    \node[state, initial] (q1) {$1$};
    \node[state] at (4, -1) (q2) {$2$};
    \node[state] at (7, -6) (q3) {$3$};
    \node[state] at (7, -9) (q4) {$4$};
    \node[state] at (5, -12) (q5) {$5$};
    \node[state] at (0, -14) (q6) {$6$};
    \node[state] at (-4, -12) (q7) {$7$};
    \node[state] at (-7, -9) (q8) {$8$};
    \node[state] at (-7, -6) (q9) {$9$};
    \node[state] at (-4, -1) (q10) {$10$};
    
    \draw (q1) edge[bend left, above, color=red] node{C/C} (q2);
    \draw (q1) edge[bend left, above, color=blue] node{D/D} (q3);
    
    \draw (q2) edge[bend left, below, color=red] node{C/C} (q1);
    \draw (q2) edge[bend right, left, color=blue] node{D/D} (q4);
    
    \draw (q3) edge[right, color=red] node{C/C} (q4);
    \draw (q3) edge[bend right, left, color=blue] node{D/D} (q5);
    
    \draw (q4) edge[right, color=red] node{C/D} (q5);
    \draw (q4) edge[bend right, left, color=blue] node{D/C} (q6);
    
    \draw (q5) edge[loop below, below, color=red] node{C/D} (q5);
    \draw (q5) edge[bend left, above, color=blue] node{D/D} (q7);
    
    \draw (q6) edge[left, color=red] node{C/D} (q2);
    \draw (q6) edge[bend left, right, color=blue] node{D/C} (q9);
    
    \draw (q7) edge[below, color=red] node{C/D} (q5);
    \draw (q7) edge[bend left, above, color=blue] node{D/D} (q8);
    
    \draw (q8) edge[bend left, above, color=red] node{C/D} (q5);
    \draw (q8) edge[below, color=blue] node{D/C} (q5);
    
    \draw (q9) edge[bend left, above, color=red] node{C/C} (q2);
    \draw (q9) edge[bend left, above, color=blue] node{D/C} (q10);
    
    \draw (q10) edge[bend left, above, color=red] node{C/D} (q2);
    \draw (q10) edge[bend left, above, color=blue] node{D/C} (q4);
    
\end{tikzpicture}
\caption{FourthPrac Diagram}
\label{fig:5}       
\end{figure}

Over the course of the evolution from \emph{SecondPrac} to \emph{FourthPrac}, there were a number of states as well as moves that were changed. Out of the 20 possible moves, 12 of them stayed the same, for example, the first move $[1, C, 2, C]$ were the same for both strategies. The first six moves stay the exact same, as the first change in strategy comes after the Fourth state. In \emph{SecondPrac}, if the opponent Cooperates while the player is in State 4, the strategy will go to State 5 and Cooperate. In \emph{FourthPrac}, under those same conditions, the strategy will go to State 5 and Defect. In total, there were three instances, where the strategy only changed a move, from Cooperate to Defect or Defect to Cooperate. There were two instances where the state only changed. And there were three instances where both state and move were changed.

\section{Final Strategy}
From the \emph{FourthPrac} strategy, Dr. Fryer evolved it using 500 generations and a 8-state FSM was created \emph{Evolved FSM 8}. This resulted in the top score of the Tournament. 

\begin{table}[ht]
\caption{Top Strategies}\label{tab:5}

\csvreader[
  longtable=llr,
  table head= 
    \toprule\bfseries Rank &\bfseries Name &\bfseries Median Score\\ \midrule\endhead
    \bottomrule\endfoot,
  late after line=\\,
  filter= \value{csvrow} < 15,
  before reading={\catcode`\#=12},
  after reading={\catcode`\#=6}
]{csvs/prof_evolve.csv}{1=\Rank,2=\Name, 3=\Median_score}{\Rank & \Name & \Median_score}
\end{table}
{\addtocounter{table}{-1}}

\begin{figure}

\tikzset{
->, 
node distance=3cm, 
every state/.style={circle, fill=gray!10, draw=green!60, radius=1cm}, 
initial text=$C$, 
}

\begin{tikzpicture}

    \node[state] (p1) {1};
    \node[state] at (5, -3) (p2) {2};
    \node[state] at (7, -6) (p3) {3};
    \node[state] at (5, -9) (p4) {4};
    \node[state, initial] at (0, -12) (p5) {5};
    \node[state] at (-5, -9) (p6) {6};
    \node[state] at (-7, -6) (p7) {7};
    \node[state] at (-5, -3) (p8) {8};
    
    \draw (p1) edge[bend right, above, color=red] node{C/C} (p3);
    \draw (p1) edge[bend right, above, color=blue] node{D/C} (p8);
    
    \draw (p2) edge[bend right, above, color=red] node{C/D} (p1);
    \draw (p2) edge[bend left, above, color=blue] node{D/D} (p5);
    
    \draw (p3) edge[loop, above, color=red] node{C/D} (p3);
    \draw (p3) edge[bend left, above, color=blue] node{D/D} (p8);
    
    \draw (p4) edge[bend left, above, color=red] node{C/D} (p7);
    \draw (p4) edge[bend left, above, color=blue] node{D/C} (p5);
    
    \draw (p5) edge[loop above, color=red] node{C/C} (p5);
    \draw (p5) edge[bend right, above, color=blue] node{D/D} (p7);
    
    \draw (p6) edge[bend right, above, color=red] node{C/D} (p3);
    \draw (p6) edge[bend left, above, color=blue] node{D/D} (p8);
    
    \draw (p7) edge[bend left, above, color=red] node{C/C} (p4);
    \draw (p7) edge[bend right, above, color=blue] node{D/D} (p6);
    
    \draw (p8) edge[bend left, above, color=red] node{C/C} (p3);
    \draw (p8) edge[bend left, above, color=blue] node{D/D} (p4);
    
\end{tikzpicture}
\caption{FourthPrac Diagram}
\label{fig:6}       
\end{figure}
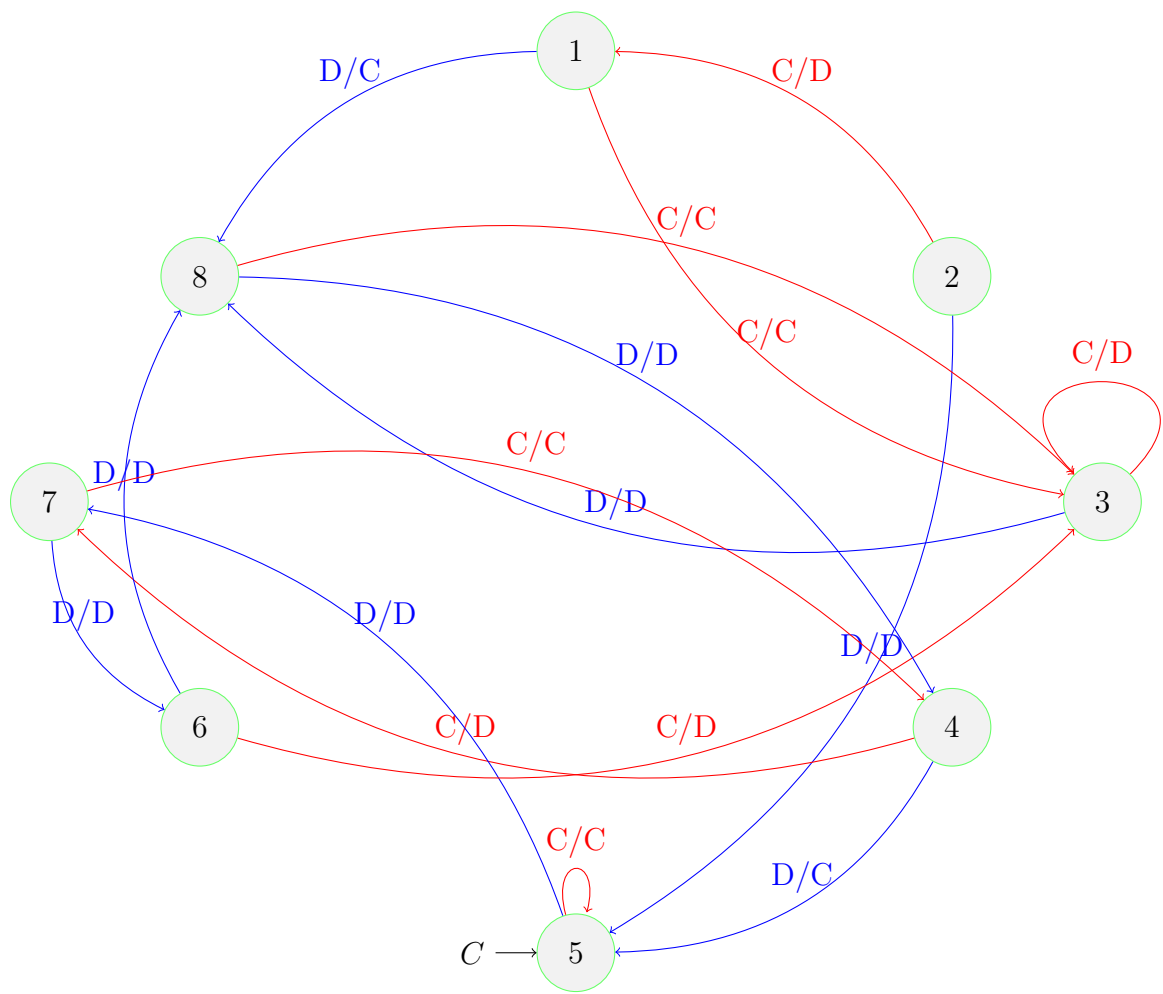

After reviewing the new strategy, 2 states out of the 8-State FSM were not accessible, so the 8-State FSM turned out to be a 6-State FSM. \emph{EvolvedFSM8} started at State 5, and looking though all the possibilities or combination of moves. State 1 and State 2 are not accessible. So those states can be removed from the diagram such that the eight state FSM, became a six state FSM, called \emph{EvolvedFSM6}. With the loss of the two states, \emph{EvolvedFSM6} still holds the top spot in the Axelrod Tournament.

\begin{table}[ht]
\caption{Top Strategies}\label{tab:6}

\csvreader[
  longtable=llr,
  table head= 
    \toprule\bfseries Rank &\bfseries Name &\bfseries Median Score\\ \midrule\endhead
    \bottomrule\endfoot,
  late after line=\\,
  filter= \value{csvrow} < 15,
  before reading={\catcode`\#=12},
  after reading={\catcode`\#=6}
]{csvs/thesis_prof_6_prac.csv}{1=\Rank,2=\Name, 3=\Median_score}{\Rank & \Name & \Median_score}
\end{table}
{\addtocounter{table}{-1}} 

Since \emph{EvolvedFSM8} starts at State 5, we will look at the possible states to get to. State 5 has two possibilities, stay in State 5 or go to State 7. There are now possible two states in the FSM. From State 7, there are two possibilities, go to State 4 or go to State 6, which leaves us with four states. The two possibilities from State 4 are to go to State 5 or go back to State 7. These are two previous states that have been identified. There are still four states. From State 6, there are two possibilities, go to State 3 or to State 8. Now there are six possible states. The possibilities for State 3 are to stay in State 3 or go to State 8. Those two possibilities are already in the FSM, so there are no new states being created. State 8 will either go to State 3 or will go back to State 4, and both of these states have been identified. This shows that there is no need to go be in State 1 or State 2 because that will never occur in this FSM since our initial state is State 5. So the 8-state FSM can be rewritten as a 6-state FSM.

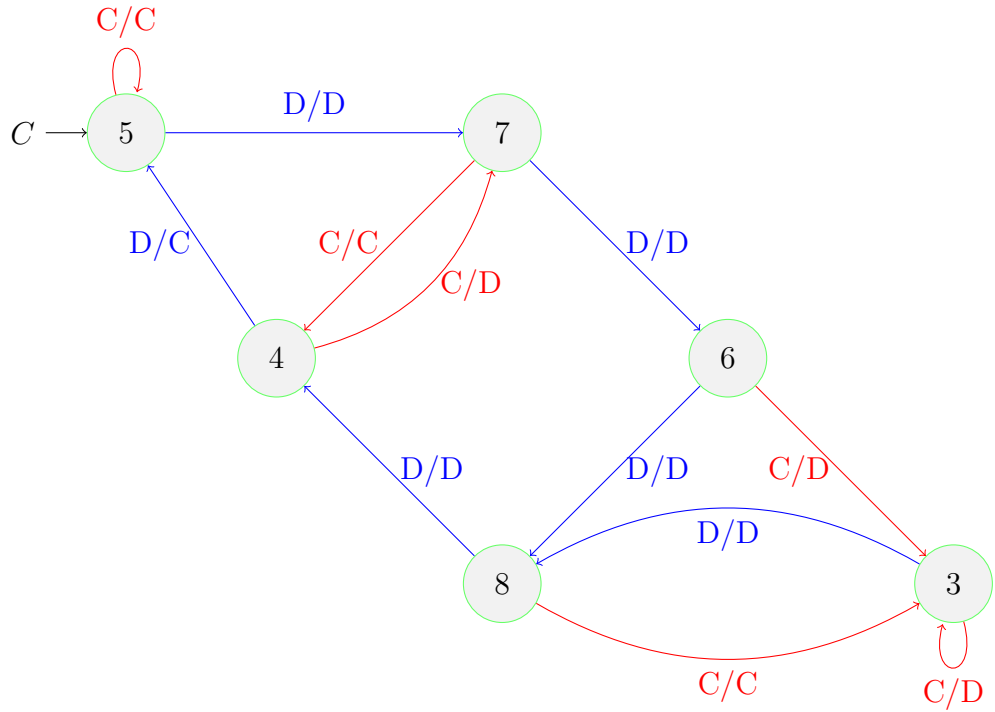
\begin{figure}

\tikzset{
->, 
node distance=3cm, 
every state/.style={circle, fill=gray!10, draw=green!60, radius=1cm}, 
initial text=$C$, 
}

\begin{tikzpicture}

    \node[state] at (6, -6) (p3) {3};
    \node[state] at (-3, -3) (p4) {4};
    \node[state, initial] at (-5, 0) (p5) {5};
    \node[state] at (3, -3) (p6) {6};
    \node[state] (p7) {7};
    \node[state] at (0, -6) (p8) {8};

    \draw (p3) edge[loop below, color=red] node{C/D} (p3);
    \draw (p3) edge[bend right, below, color=blue] node{D/D} (p8);
    
    \draw (p4) edge[bend right, right, color=red] node{C/D} (p7);
    \draw (p4) edge[left, color=blue] node{D/C} (p5);
    
    \draw (p5) edge[loop above, color=red] node{C/C} (p5);
    \draw (p5) edge[above, color=blue] node{D/D} (p7);
    
    \draw (p6) edge[left, color=red] node{C/D} (p3);
    \draw (p6) edge[right, color=blue] node{D/D} (p8);
    
    \draw (p7) edge[left, color=red] node{C/C} (p4);
    \draw (p7) edge[right, color=blue] node{D/D} (p6);
    
    \draw (p8) edge[bend right, below, color=red] node{C/C} (p3);
    \draw (p8) edge[right, color=blue] node{D/D} (p4);
\end{tikzpicture}
\caption{EvolvedFSM8 to EvolvedFSM6}
\label{fig:7}       
\end{figure}

We have done an initial dive into looking at when our evolved strategy became the top strategy in the Axelrod Tournament. It seemed to occur after the $351^{st}$ evolutionary generation. The $351^{st}$ evolutionary generation is also important because it is when the strategy went from a 7-state FSM to a 6-state FSM. On the parameter file, where each evolution generation is stored, there was a significant jump in the $351^{st}$ generation. In Generation 350, the average score was $2.871814908$, and on Generation 351, the average score was $2.885947477$. This was a difference in $0.014132569$ points, which is significant. For context, there were over 200 generations, where the score stayed with $2.87$.

When running Generation 350 and Generation 351 against the Short Run Strategies in the Axelrod Tournament, they both place high. Generation 350 places $3^{rd}$, while Generation 351 takes the top score. It must be at this specific iteration where the score becomes the highest.

\section{Long Run Strategies}
From all the previous tournaments we have run using the Axelrod Library, we mainly focused on the Short Run Time strategies. One of the main reasons for doing that was my computer was not capable of running a tournament that included the Long Run Time strategies. We assumed that our strategy had the highest score, but we could not know for sure until it ran against the Full Tournament library, which included the Long Run Time Strategies. Table~\ref{tab:8} shows the Full Tournament which includes the Long Run Time Strategies before adding \emph{EvolvedFSM6}.

Dr. Fryer ran the \emph{EvolvedFSM6} on his computer to generate these results for our strategy in the Axelrod Tournament. Table~\ref{tab:9} refers to the Tournament with all the strategies, including the Long Run Time Strategies in the Axelrod Library. \emph{EvolvedFSM6} will be the top score against all the different strategies in the Axelrod Tournament.

\section{Conclusion}
We have shown that a six state Finite State Machine can be the top score in the Axelrod Tournament. By initially having a strategy that tries to follow the Cooperation Rates of the Top Strategies, a strategy can place fairly high in the tournament. The strategy can then be evolved in order to maximize a top score.

With further research and time, we want to be able to increase our current score to see what is the highest score possible in the Axelrod Tournament can be. Starting off with a high score prior to evolving the score played a major role in advancing our strategy. 

\begin{table}[p]
\caption{Top Strategies}\label{tab:7}

\csvreader[
  longtable=llr,
  table head= 
    \toprule\bfseries Rank &\bfseries Name &\bfseries Median Score\\ \midrule\endhead
    \bottomrule\endfoot,
  late after line=\\,
  filter= \value{csvrow} < 15,
  before reading={\catcode`\#=12},
  after reading={\catcode`\#=6}
]{csvs/profparams.csv}{1=\Rank,2=\Name, 3=\Median_score}{\Rank & \Name & \Median_score}
\end{table}
{\addtocounter{table}{-1}} 

\begin{table}[p]
\caption{Long Run Strategies}\label{tab:8}
%
\csvreader[
  longtable=llr,
  table head= 
    \toprule\bfseries Rank &\bfseries Name &\bfseries Median Score\\ \midrule\endhead
    \bottomrule\endfoot,
  late after line=\\,
  filter= \value{csvrow} < 15,
  before reading={\catcode`\#=12},
  after reading={\catcode`\#=6}
]{csvs/pre.csv}{1=\Rank,2=\Name, 3=\Median_score}{\Rank & \Name & \Median_score}
\end{table}
{\addtocounter{table}{-1}} 

\begin{table}[p]
\caption{Long Run Strategies}\label{tab:9}
%
\csvreader[
  longtable=llr,
  table head= 
    \toprule\bfseries Rank &\bfseries Name &\bfseries Median Score\\ \midrule\endhead
    \bottomrule\endfoot,
  late after line=\\,
  filter= \value{csvrow} < 15,
  before reading={\catcode`\#=12},
  after reading={\catcode`\#=6}
]{csvs/post.csv}{1=\Rank,2=\Name, 3=\Median_score}{\Rank & \Name & \Median_score}
\end{table}
{\addtocounter{table}{-1}} 

\bibliographystyle{abbrv}
\bibliography{name.bib}


\end{document}